\documentclass[prl,twocolumn,superscriptaddress,showpacs,nobalancelastpage]{revtex4}
\usepackage{graphicx}

\begin{document}

\title{Electrostatic confinement of electrons in graphene nano-ribbons}

\author{Xinglan Liu}
\affiliation{Kavli Institute of NanoScience, Delft University of Technology, PO Box 5046, 2600 GA, The Netherlands}

\author{Jeroen B. Oostinga}
\affiliation{Kavli Institute of NanoScience, Delft University of Technology, PO Box 5046, 2600 GA, The Netherlands}
\affiliation{DPMC and GAP, University of Geneva, 24 quai Ernest Ansermet, Geneva CH1211, Switzerland}

\author{Alberto F. Morpurgo}
\affiliation{DPMC and GAP, University of Geneva, 24 quai Ernest Ansermet, Geneva CH1211, Switzerland}

\author{Lieven M.K. Vandersypen}
\affiliation{Kavli Institute of NanoScience, Delft University of Technology, PO Box 5046, 2600 GA, The Netherlands}
\email{l.m.k.vandersypen@tudelft.nl}

\date{\today}

\begin{abstract}

Coulomb blockade is observed in a graphene nanoribbon device with a top gate. When two $pn$-junctions are formed via the back gate and the local top gate, electrons are confined between the $pn$-junctions which act as the barriers. When no $pn$-junctions are induced by the gate voltages, electrons are still confined, as a result of strong disorder, but in a larger area. Measurements on five other devices with different dimensions yield consistent results.

\end{abstract}

\pacs{85.35.p, 73.23.-b, 72.80.Rj, 73.20.Fz}
%85.35.p  	Nanoelectronic devices
%73.23.-b   Electronic transport in mesoscopic systems
%72.80.Rj   Fullerenes and related materials
%73.20.Fz 	Weak or Anderson localization
%81.05.Uw 	Carbon, diamond, graphite

\maketitle

Confinement of the Dirac particles is of particular importance for the realization of nano-electronic devices in graphene such as quantum dots \cite{Leo}. These would enable one to perform single-level spectroscopy of Dirac particles, study their spin and valley degrees of freedom, and explore their potential for quantum coherent control \cite{Bjorn}. In conventional semiconductors, particles can be confined by potential barriers created via electrostatic gates. This approach permits independent control of the number of electrons on the island, the tunnel coupling between the island and the reservoirs, as well as the tunnel coupling between neighbouring islands. Such flexibility and versatility has been instrumental for a wide variety of mesoscopics experiments. In graphene, this approach normally fails, due to the absence of a bandgap and the presence of Klein tunnelling \cite{Klein,GeimNM}. In previous studies, graphene has been etched into small islands, separated from the reservoir by narrow constrictions \cite{Ponom,ChrisAPL,Johannes}, but here it is difficult to tune the barriers. Alternatively, a bandgap could be created in graphene first, so that electrostatic gates can again be used for confinement. Theoretically, a bandgap is predicted in graphene nanoribbons (GNRs) due mainly to quantum confinement \cite{Nakada,BreyFertig,SonPRL}. Experimentally, a transport gap has indeed been observed in GNR devices \cite{Han,HJDSci,Avouris,ChrisNew}, but its origin is still under debate.

Here we experimentally investigate GNR devices with a local top gate (TG) and a global back gate (BG) where the transport gap in the GNR enables electrostatic confinement by the gates. Electrons are confined in an island where the barriers are formed by the $pn$-junctions induced at the two edges of the TG, as demonstrated by the capacitances analysis of the measured Coulomb blockade. On the other hand, when no $pn$-junctions were induced by the gates, Coulomb blockade was also observed, showing a larger confinement area. Here the island may be due to Anderson localization. Consistent results were found in five other devices with different dimensions.

\begin{figure}[!t]
\includegraphics[width=3.4in]{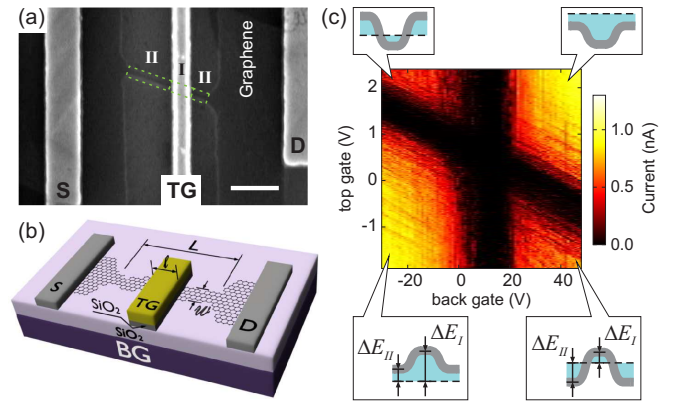}
\caption{(Color online) (a) Scanning electron microscope image of device $E$ (scale bar 300 nm). The green dashed lines indicate segment I and II discussed in the text. (b) Schematic drawing of the device. (c) Current as a function of the back gate and top gate voltages for device $A$ at $V_{bias} = 150~\mu$V and $T = 4$ K. The insets illustrate the potential landscapes that are created in the four corners of the plot. They represent energy diagrams along the ribbon length, where the gray band indicates the transport gap and the dashed lines represent the Fermi level $E_F$. $\Delta E_{I(II)}$ indicates the doping level in segment I (II).}
\label{devices}
\end{figure}

Six devices ($A$ to $F$) are fabricated on graphene flakes deposited on a substrate by mechanical exfoliation of natural graphite \cite{Novos05}. The substrate consists of highly $p$-doped silicon, acting as a back gate (BG), capped by 285 nm of SiO$_2$. From their optical contrast against the substrate, we estimate that the flakes are single-layer \cite{Jeroen}. Three electron beam lithography steps were used for patterning the devices. First, selected graphene flakes are patterned into GNRs, using PMMA as an etching mask and an Ar plasma for etching (for device $F$ an O$_2$ plasma was used). Next we pattern a single top gate across each ribbon. The TG consists of 10/5/40 nm thick evaporated SiO$_2$/Ti/Au, and it covers only part of the ribbon, denoted as segment I (see Fig. 1a). The remainder of the ribbon (segments II) connects to wider pieces of graphene, which are contacted by 10/40 nm thick Ti/Au source (S) and drain (D) electrodes. The device is schematically illustrated in Fig. 1b, and the relevant device dimensions are given in Table \ref{lieven}.

\begin{table}
\begin{tabular*}{0.45\textwidth}{@{\extracolsep{\fill}} l|c|cccccc}
                & Label & $A$ & $B$ & $C$ & $D$ & $E$ & $F$ \\
  \hline
  GNR width & $w$ (nm) & 60  & 50 & 50 & 50 & 40 & 40 \\
  GNR length & $L$ (nm) & 2000 & 1500 & 1000 & 700 & 520 &520 \\
  TG width & $l$ (nm) & 500 & 400 & 200 & 140 & 100 & 50 \\
\end{tabular*}
\caption{The dimensions of devices $A$ to $F$. The TG dielectric is $d=10$ nm thick SiO$_2$ for all devices.}
\label{lieven}
\end{table}

All measurements were performed in a $^3$He system at a base temperature of 350 mK, unless stated otherwise. We measured the two terminal resistance through the top gated GNR devices by applying a DC voltage bias, $V_{bias}$, on the source electrode and measuring the current at the drain electrode.

By tuning the BG and TG voltages, we can shape the potential landscape along the ribbon. Fig. 1c shows the low bias conductance of device $A$ as a function of $V_{BG}$ and $V_{TG}$ at $T=4$ K. Along the dark vertical band, the conductance is suppressed as $E_F$ is within the transport gap in segment II. Along the dark diagonal band, the TG and BG dope the graphene with opposite polarity and $E_F$ lies in the transport gap in segment I. At zero gate voltages, the device was unintentionally hole doped.

In the lower right (upper left) corner of Fig. 1c, the ribbon is in a $npn$ ($pnp$) configuration. In this regime, holes (electrons) can be confined in the area (segment I) between the two $pn$-junctions owing to the presence of the transport gap \cite{Efetov,Gary}. We thus expect Coulomb blockade in the $npn$ and $pnp$ regimes. In the lower left (upper right) corner, the ribbon is in a $pp'p$ ($nn'n$) configuration. Here Fabry-Perot type resonances could occur between the two steps in the potential landscape, but no Coulomb blockade is expected in an ideal ribbon, as there are no barriers. The difference in energy from $E_F$ to the middle of the transport gap in segment I and II are denoted as $\Delta E_{I}$ and $\Delta E_{II}$, respectively, which is a measure of the doping level.

% device B npn
In the $npn$ configuration, we observe pronounced current peaks separated by zero-current regions as $V_{TG}$ is swept. A representative measurement is shown in Fig. 2d for device $B$, measured in the gate voltage configuration indicated by the black arrows in Fig. 2a and 3b. High bias measurements in the same range (Fig. 2e) show diamond-shaped regions in the $V_{bias}-V_{TG}$ plane, in which current is blocked. Both are characteristic of Coulomb blockade due to the formation of an island that is only weakly coupled to the leads.

In this device, over 700 Coulomb peaks were resolved in the range $-2$ V$<V_{TG}<-0.2$ V, corresponding to a large change in doping level in segment I, $-360$ meV$ \lesssim \Delta E_I\lesssim -240$ meV (Fig. 3a). Here $\Delta E_I$ reflects the doping level in segment I (Fig. 1c inset), and is roughly estimated by considering the density of states of bulk graphene, $\Delta E_I = (\pm)\hbar v_F \sqrt{\pi n_I}$, where $n_I$ is the carrier density in region I, $v_F$ is the Fermi velocity of bulk graphene, and the $(+)$ and $(-)$ signs represent electron and hole doping, respectively. The spacings between neighbouring peaks, $\Delta V_{TG}^{npn}$, are shown in Fig. 3a (black spheres), as a function of the peak positions. The average value $\langle \Delta V_{TG}^{npn}\rangle=2.0\pm 0.4$ mV corresponds to a TG capacitance $C_{TG}=70-100$ aF, close to what one would expect from simple parallel plate capacitance between the TG and segment I, $C_{TG}^{\parallel} = \epsilon_0 \epsilon_r w l/d = 70$ aF, where $\epsilon_r = 3.9$ is the relative permittivity of SiO$_2$. In addition, the capacitance to the back gate, measured to be $\sim 3.9$ aF, compares well to the value expected from the geometry of an island of area $(wl)$. The agreement demonstrates that for this device, an island is formed between the two $pn$-junctions in the $npn$ configuration.

In addition, we measured over 100 Coulomb diamonds similar to Fig. 2e, and the extracted addition energy $E_a^{npn}$ for each diamond is shown in Fig. 3a with green triangles. The average addition energy is $\langle E_a^{npn} \rangle=1.0 \pm 0.4$ meV. From Fig. 3a, no shell filling or evident top gate voltage dependence is observed in either $\Delta V_{TG}$ or $E_a$, but both quantities show a large spread similar to \cite{ChrisAPL,Ponom}, due to contributions from both the level spacing and strong disorder from the ribbon edges, which is discussed further below. In all these measurements, segments II of the GNR were heavily $n$-doped such that disorder in the leads was largely screened ($V_{BG}=+81$ V) \cite{gap}.

% device B ppp
Unexpectedly, Coulomb blockade was also observed when no $pn$-junctions are present. Fig. 2b and 2c show a representative current trace and Coulomb diamonds measured from the same device in a $pp'p$ configuration indicated by the red arrows in Fig. 2a and 3d. Over 1400 current peaks were observed in the range $-1$ V$<V_{TG}<2.2$ V, corresponding to a change in doping level $\Delta E_I$ from $-430$ meV to $ -230$ meV (Fig. 3c). We measured 70 Coulomb diamonds in the same range. The extracted peak spacings $\Delta V_{TG}^{pp'p}$ and addition energy $E_a^{pp'p}$ are shown in Fig. 3c with red spheres and blue triangles, respectively. The measurements were taken at $V_{BG}=0$ (red arrow in Fig. 2a), such that segment II of the GNR was heavily $p$-doped.

The average peak spacing in the $pp'p$ configuration $\langle \Delta V_{TG}^{pp'p}\rangle=1.8 \pm 0.4$ mV is very close to $\langle \Delta V_{TG}^{npn}\rangle$, but the average addition energy $\langle E_a^{pp'p} \rangle=0.5 \pm 0.2$ meV is only half the value of $\langle E_a^{npn} \rangle$. The back gate capacitance in $pp'p$ is $\sim 10$ aF, indicating an island area of $50$ nm by $700$ nm, larger than $(wl)$ (assuming the island extends over the entire ribbon width in the transverse direction). All average quantities were reproducible over multiple thermal cycles. Therefore the island formed in the $pp'p$ configuration is located in part under the TG, but extends to a larger area than the island in the $npn$ case.

\begin{figure}[!t]
\includegraphics[width=3.4in]{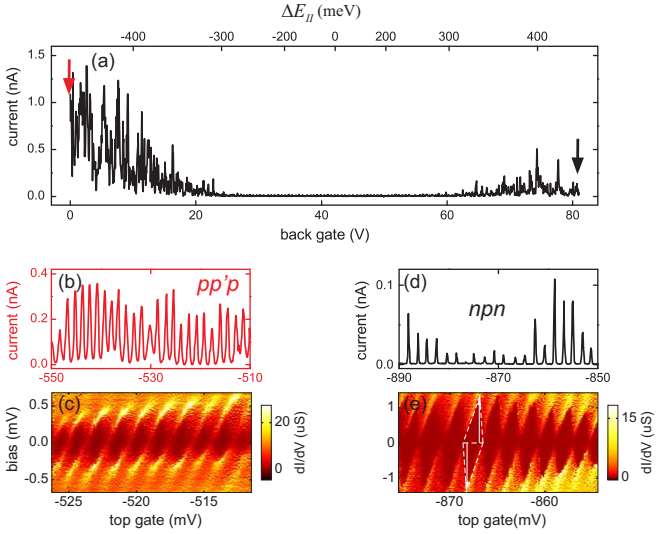}
\caption{(Color online) (a) Current as a function of $V_{BG}$ for device $B$ at $V_{TG}=-1.2$ V and $V_{bias} = 200~\mu$V. The top axis indicates the corresponding doping level in segment II, $\Delta E_{II}$, estimated in the same way as $\Delta E_I$ (see text). (b) Coulomb oscillations as a function of $V_{TG}$ in the $pp'p$ configuration. $V_{BG}=0$ as indicated by the red (gray) arrow in (a) and $V_{bias} = 100~\mu$V. (c) Differential conductance $dI/dV$ as a function of $V_{TG}$ and $V_{bias}$ (Coulomb diamonds), measured in the same $pp'p$ regime as (b). (d) Coulomb oscillations as a function of $V_{TG}$ in the $npn$ configuration. $V_{BG}=+81$ V as indicated by the black arrow in (a), and $V_{bias} = 100~\mu$V. (e) Coulomb diamonds measured in the same $npn$ regime as (d). The addition energy $E_a$ is taken as the average of the two blue arrows.}
\label{CBO_DMD}
\end{figure}

\begin{figure}[t]
\includegraphics[width=3.4in]{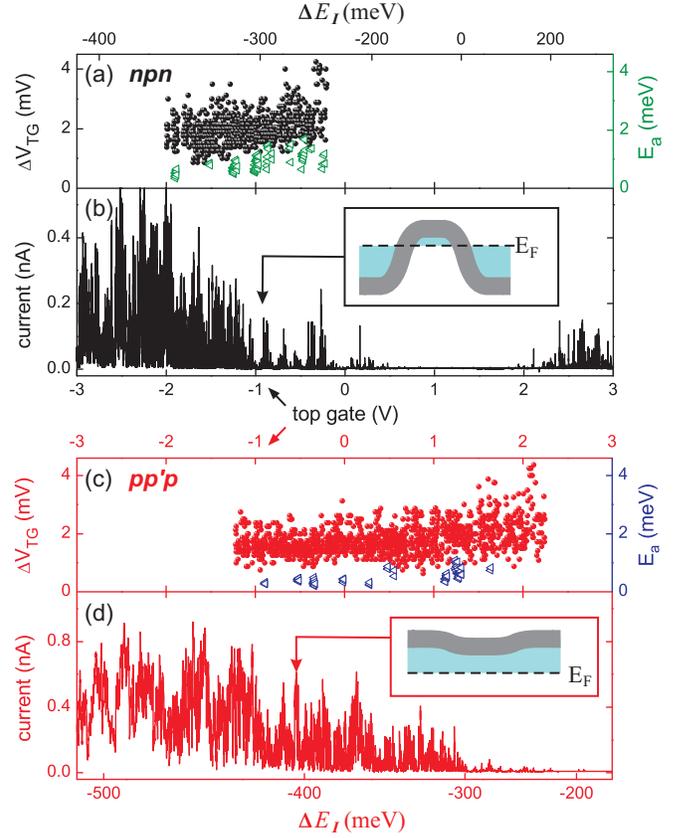}
\caption{(Color online) (a) $\Delta V_{TG}$ (black spheres) and $E_a$ (green triangles) as a function of peak positions in the $npn$ regime for device $B$. $V_{BG}=+81$ V. (b) Current (black line) as a function of $V_{TG}$ at $V_{bias}=100~\mu$V and $V_{BG}=+81$ V. (c) $\Delta V_{TG}$ (red spheres) and $E_a$ (blue triangles) in the $pp'p$ regime for at $V_{BG}=0$. (d) Current (red line) as a function of $V_{TG}$ at $V_{bias}=100~\mu$V and $V_{BG}=0$. The corresponding doping level in segment I, $\Delta E_I$, are indicated in the top axis (a) and bottom axis (d). The two insets illustrate the potential landscapes created at the gate voltage configurations where Fig. 2b and 2d are taken.}
\label{scatt_vertical}
\end{figure}

% 3 GENERALIZE NPN
Coulomb blockade in the $npn$ configuration was found in three other devices of different dimensions ($A, C, E$) where the $npn$ regime could be accessed (the various devices exhibited different positions of the charge neutrality point). For device $E$, also the $pnp$ configuration could be reached, where the measurement results are analogous to those for $npn$. The extracted $\langle \Delta V_{TG}^{npn} \rangle$ and $\langle E_a^{npn} \rangle$ for these devices are summarized in Fig. 4a (black filled circles) and 4b (black filled squares), respectively. For all four devices, the measured $\langle \Delta V_{TG}^{npn} \rangle$ agrees quantitatively well with the numerically computed $e/C_{TG}$ \cite{Poisson} by considering an island of size $wl$ (black dotted line in Fig. 4a). Moreover, both $\langle \Delta V_{TG}^{npn} \rangle$ and $\langle E_a^{npn} \rangle$ increase with decreasing area $(w l)$, consistent with the observation in device $B$, i.e. carriers are confined in segment I in the $npn$ configuration.

% 4 GENERALIZE PPP
Furthermore, we observed Coulomb blockade in the $pp'p$ regime in all six devices $A$ to $F$. For devices $A, B, E$, we could also access the $nn'n$ configuration and the results are similar to those in the $pp'p$ case. For all devices, the average addition energy, $\langle E_a^{pp'p}\rangle$ in the $pp'p$ regime is much smaller than that in the $npn$ case, and does not vary much despite the differences in device dimensions (except for device $F$), as shown by the red open squares in Fig. 4b (no clear Coulomb diamonds were observed in device $A$ in the $pp'p$ regime). In devices $B$ and $D$, the measured back gate capacitance indicates that the island sizes are $50\times700$ nm$^2$ and $50\times250$ nm$^2$, respectively (Fig. 4b inset), significantly larger than the area of segment I. The capacitance of this large island to the relative narrow TG, $C_{TG}^{pp'p}$, is still roughly the same as that for an island limited to segment I, so the peak spacings in $pp'p$ (Fig. 4a, red circles) are similar to those for $npn$.

The reproducible scaling of addition energy and peak spacing as a function of devices dimensions is consistent with the results obtained from device $B$: when
$pn$-junctions are induced by the TG and BG, an island is formed in between the junctions; without the $pn$-junctions, a much larger island is formed, presumably due to disorder. However, the source and drain capacitances are comparable ($npn$) or even larger ($pp'p$) than $C_{TG}$, and contribute more than half of the total capacitance. This means that extracting the island size from $E_a$ may be unreliable. For device $B$ ($npn$ and $pp'p$) and $D$ ($pp'p$), we have measured the back gate capacitance, which gives an independent estimate of the island size (Fig. 4b inset) and is in agreement with our interpretation of the island size in the two regimes.

If islands are induced by disorder in the $pp'p$ case, there are likely to be disorder-induced islands in the $npn$ case as well in addition to the islands formed by the $pn$-junctions. Indeed evidence of multiple islands was observed experimentally in several cases (not shown). The presence of the additional islands contributes to the large spread in peak spacings and addition energies mentioned earlier.

These disorder-induced islands in the $pp'p$ regimes are 5-10 times longer than the ribbon width, which could be explained by Anderson localization, due to strong scattering at the rough ribbon edges as proposed in \cite{White,YoonGuo,Evaldsson,Lewenkopf,Basu}. This reveals a different aspect of the electronic properties of the GNRs compared to other work, where the extent of the island is found to be comparable with the ribbon width \cite{ChrisNew,Sols,DG2}. Further studies are needed in order to clarify the underlying mechanisms behind the various observations.

\begin{figure}[!t]
\includegraphics[width=3.4in]{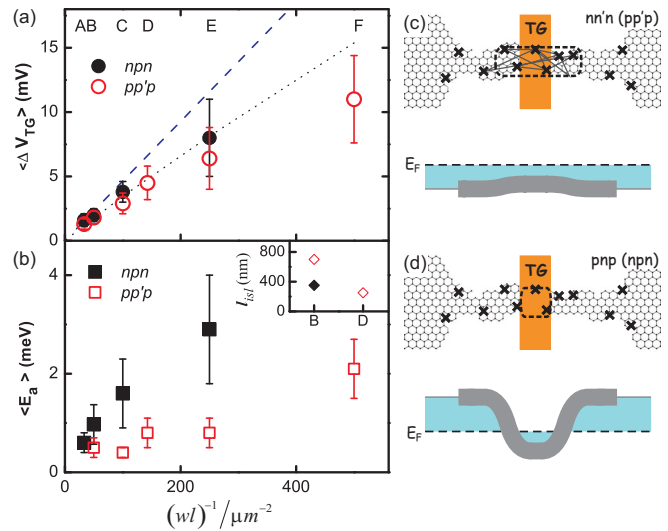}
\caption{(Color online) (a) Average peak spacing versus the inverse area of segment I, $(wl)^{-1}$, in the $npn$ (black filled circles) and $pp'p$ (red open circles) configurations, for devices $A$ through $F$ as indicated. The blue dashed line is the peak spacing estimated by $e/C_{TG}^{\parallel}$, where $C_{TG}^{\parallel}$ is a parallel plate capacitance between the TG and GNR (see text). The values of $e/C_{TG}$ were also computed numerically by considering an island of area $(w l)$ (black dotted line) \cite{Poisson}. (b) Average addition energy extracted from Coulomb diamonds measured in the $npn$ (black filled squares) and $pp'p$ (red open squares) configurations versus $(wl)^{-1}$. The error bars in (a,b) represent standard deviation. Inset: the island size in the direction along the ribbon length ($l_{isl}$) for devices $B$ and $D$ in both $npn$ (black filled diamonds) and $pp'p$ (red open diamonds) regimes extracted from the back gate capacitances. (c,d) Schematic illustrations of the device, with the $\times$ symbol representing the scattering sites and the island enclosed in dashed lines. The underlying energy diagrams for the $nn'n$ (c) and $pnp$ (d) regimes are also shown where the gray band represents the transport gap in the GNRs.}
\label{summary}
\end{figure}

In conclusion, a single electron transistor is formed in graphene nanoribbon devices with single top gates. Two $pn$-junctions at the two edges of the top gate induced by the top gate and back gate voltages act as barriers to form an island. Hundreds of Coulomb peaks were observed in this regime. In the absence of the $pn$-junctions, regular Coulomb blockade is also observed where the island can be induced by ribbon edge disorder. Observations from measurements of five other devices give consistent results. We anticipate that multiple top gates on a graphene nanoribbon will offer additional control for future device applications, and provide further insight into the electronic properties of graphene nanoribbons.

\begin{acknowledgments}

We thank K. Ensslin, M. Fogler, D. Goldhaber-Gordon, P. Kim, Y. Nazarov, C. Stampfer, G. Steele for useful discussions, and L.P. Kouwenhoven for the use of a $^3$He system. This work was supported by the Dutch Foundation for Fundamental Research on Matter (FOM).

\end{acknowledgments}


\begin{references}
\bibitem{gap}
When the carrier density in segment I or II is low, we typically observe that low bias conductance vanishes and Coulomb diamonds overlap.
\end{references}
\end{document}